\documentclass[10pt]{iopart}
\usepackage{iopams}
\usepackage {graphicx}

\begin{document}
\title[]{The LTP Experiment on the LISA Pathfinder Mission}
\author{Anza S$^{1}$, Armano M$^{2}$, Balaguer E$^{3}$, Benedetti
M$^{4}$, Boatella C$^{1}$, Bosetti P$^{4}$, Bortoluzzi D$^{4}$,
Brandt N$^{5}$, Braxmaier C$^{5}$, Caldwell M$^{6}$, Carbone
L$^{2}$, Cavalleri A$^{2}$, Ciccolella A$^{3}$, Cristofolini
I$^{4}$, Cruise M$^{7}$, Da Lio M$^{4}$, Danzmann K$^{8}$,
Desiderio D$^{9}$, Dolesi R$^{2}$, Dunbar N$^{10}$, Fichter
W$^{5}$, Garcia C$^{3}$, Garcia-Berro E$^{11}$, Garcia Marin A
F$^{8}$, Gerndt R$^{5}$, Gianolio A$^{3}$, Giardini D$^{12}$,
Gruenagel R$^{3}$, Hammesfahr A$^{5}$, Heinzel G$^{5}$, Hough
J$^{13}$, Hoyland D$^{7}$, Hueller M$^{2}$, Jennrich O$^{3}$,
Johann U$^{5}$, Kemble S$^{10}$, Killow C$^{13}$, Kolbe D$^{5}$,
Landgraf M$^{14}$, Lobo A$^{15}$, Lorizzo V$^{9}$, Mance D$^{12}$,
Middleton K$^{6}$, Nappo F$^{9}$, Nofrarias M$^{1}$, Racca
G$^{3}$, Ramos J$^{11}$, Robertson D$^{13}$, Sallusti M$^{3}$,
Sandford M$^{6}$, Sanjuan J$^{1}$, Sarra P$^{9}$, Selig A$^{16}$,
Shaul D$^{17}$, Smart D$^{6}$, Smit M$^{16}$, Stagnaro L$^{3}$,
Sumner T$^{17}$, Tirabassi C$^{3}$, Tobin S$^{6}$, Vitale
S$^{2}$\footnote{Corresponding author: Stefano.Vitale@unitn.it},
Wand V$^{8}$, Ward H$^{13}$, Weber W J$^{2}$, Zweifel P$^{12}$}

\address{$^1$ Institut d'Estuds Espacials de Catalunya, Barcelona,
Spain}
\address{$^2$ Department of Physics and INFN, University of Trento, 38050 Povo (TN),
Italy}
\address{$^3$ ESA-ESTEC, 2200 AG Noordwijk (The Netherlands)}
\address{$^4$ Department of Mechanical and Structural Engineering, University of
Trento, 38050 Trento, Italy}
\address{$^5$ EADS Astrium GmbH, Friedrichshafen, Immenstaad
88090,~Germany}
\address{$^6$ Rutherford Appleton Laboratory, Chilton-Didcot, UK}
\address{$^7$ Department of Physics {\&} Astronomy, University of Birmingham,
UK}
\address{$^8$ Max-Planck-Institut f\"{u}r Gravitationsphysik
(Albert-Einstein-Institut) and Universit\"{a}t Hannover, Hannover,
Germany}
\address{$^9$ Carlo Gavazzi Space, 20151 Milano, Italy}
\address{$^{10}$ EADS Astrium Ltd, Stevenage, Hertfordshire, SG1 2AS,
UK}
\address{$^{11}$ Univertitat Politecnica de Catalunya, Barcelona,
Spain}
\address{$^{12}$ Swiss Federal Institute of Technology Zurich, Geophysics, CH-8093
Z\"{u}rich}
\address{$^{13}$ Department of Physics and Astronomy, University of
Glasgow, Glasgow, UK}
\address{$^{14}$ ESA/ESOC, 64293 Darmstadt, Germany}
\address{$^{15}$ Universidad de Barcelona, Barcelona 08028 Spain}
\address{$^{16}$ SRON National Institute for Space Research, 3584 CA
Utrecht, the Netherlands}
\address{$^{17}$ The Blackett Laboratory, Imperial College of Science, Technology {\&}
Medicine, London, UK}

\begin{abstract}
We report on the development of the LISA Technology Package (LTP)
experiment that will fly on board the LISA Pathfinder mission of
the European Space Agency in 2008. We first summarize the science
rationale of the experiment aimed at showing the operational
feasibility of the so called Transverse{-}Traceless coordinate
frame within the accuracy needed for LISA. We then show briefly
the basic features of the instrument and we finally discuss its
projected sensitivity and the extrapolation of its results to
LISA.
\end{abstract}

\section{Introduction}
\label{sec:introduction}

Our very concept of the detection of gravitational wave by an
interferometric detector like LISA \cite{re01,re02} is based on
the operative possibility of realizing a Transverse and Traceless
(TT) coordinate frame \cite{re03}.

In this kind of coordinate frame, despite the presence of the
ripple in space{-}time curvature due to the gravitational wave, a
free particle initially at rest remains at rest, i.e. its space
coordinates do not change in time, and the proper time of a clock
sitting on the particle coincides with the coordinate time.

Despite the co-moving nature of such a frame, the distances among
particles at rest change in time because of the change of the
metric tensor, and this time variation can be detected by the
laser interferometer.

Indeed a laser beam travelling back and forth between two such
particles along an axis {\bf x} normal to the direction {\bf z} of
the gravitational wave propagation, is subject to a phase shift
$\delta \theta $(t) whose time derivative is given ${\delta \nu }
\mathord{\left/ {\vphantom {{\delta \nu } {\nu _o }}} \right.
\kern-\nulldelimiterspace} {\nu _o }$ by \cite{re03}:

\begin{equation}\label{eqn1}
\frac{d\delta \theta }{dt} = \frac{\pi c}{\lambda }\left[ {h_ +
\left( {t - \frac{2L}{c}} \right) - h_ + \left( t \right)}
\right].
\end{equation}

Here $h_ + $ is the usual definition \cite{re03} for the amplitude
of the wave, $L$ is the distance between the particles and
$\lambda $ is the wavelength of the laser. Furthermore the {\bf
x}-axis has been used to define the wave polarization, so that the
phase shift is only contributed by $h_ + $, and $t$ is the time at
which light is collected and the frequency shift is measured.

As all this holds within a linearized theory, small effects
superimpose and harmonic analysis can be applied. As a consequence
secular gravitational effects at frequencies much lower than the
observation bandwidth ($f < 10^{ - 4}$ Hz) do not matter. A TT
coordinate frame may then in principle be defined just for the
frequencies of relevance, letting the particles used to mark the
frame to change their coordinates at lower frequencies because of
their motion within the gravitational field of the Solar System.

If the particles are not at rest in the TT frame, i.e. if their
space coordinates change in time, then obviously their distances
will change also because of this motion. If the particles still
move slowly relative to light, their relative motion does not
affect the TT construction but competes with the signal in
\eref{eqn1} by providing a phase shift:
\begin{equation}\label{eqn2}
\delta \theta \left( t \right) = \frac{2\pi }{\lambda }\left\{
{x_1 \left( t \right) + x_1 \left( {t - \frac{2L}{c}} \right) -
2x_2 \left( {t - \frac{L}{c}} \right)} \right\}
\end{equation}
\noindent where $x_{1}$ is the coordinate of the particle sending
and collecting the laser beam, while $x_{2}$ is that of the
particle reflecting the light. Here coordinates are components
along the laser beam and the phase shift is calculated to first
order in $v/c$.

At measurement frequencies much lower than $c/L$ \eref{eqn2} gives
obviously
\begin{equation}\label{eqn3}
\delta \theta \left( t \right) \approx \frac{4\pi }{\lambda
}\Delta L\left( t \right) \mbox{ with } \Delta L\left( t \right) =
x_1 \left( t \right) - x_2 \left( t \right).
\end{equation}

If all coordinates may be assumed as joint stationary random
processes, the phase shift in \eref{eqn3} has a Power Spectral
Density (PSD)
\begin{eqnarray}\label{eqn4}
S_{\delta \theta } \left( \omega \right) &=& \frac{16\pi
^2}{\lambda ^2} \left\{ S_{\Delta L} \left( \omega \right) \left[
{1 - 2\sin^2 \left( \frac{\omega L}{2c} \right) } \right]
\right. +\nonumber \\
 &&\left. + 8\sin^2\left( {\frac{\omega L}{2c}} \right) \left[
{S_{x_2 } \left( \omega \right) - \cos\left( {\frac{\omega L}{c}}
\right) S_{x_1 } \left( \omega \right)} \right] \right\} \nonumber \\
&\approx& \frac{16\pi ^2}{\lambda ^2}S_{\Delta L} \left( \omega
\right)
\end{eqnarray}
\noindent where $S_{\Delta L} \left( \omega \right),\,S_{x_2 }
\left( \omega \right)$, and $S_{x_1 } \left( \omega \right) $ are
the PSD of the related quantities at angular frequency $\omega$
and the rightmost approximate equality holds for ${\left| \omega
\right|L} \mathord{\left/ {\vphantom {{\left| \omega \right|L} {c
\ll 1}}} \right. \kern-\nulldelimiterspace} {c \ll 1}$.

In a TT frame, and at low velocities, the motion of proof{-}masses
can only be caused by forces that are not due to the gravitational
wave. \Eref{eqn4} then becomes:
\begin{equation}\label{eqn5}
S_{\delta \theta } \approx \frac{16\pi ^2}{\lambda
^2}\frac{S_{\Delta F} }{\omega ^4m^2}
\end{equation}
\noindent where $S_{\Delta F} $ is the spectral density of the
\textit{difference} of force between the proof-masses.

Thus to show that a TT system can indeed be constructed with free
orbiting particles, one needs to preliminarily show that
non{-}gravitational forces on proof{-}masses, or even locally
generated gravitational forces, can be suppressed to the required
accuracy in the measurement bandwidth.

The interferometer measurement noise will also compete with the
gravitational signal in \eref{eqn1}. This noise is usually
expressed as an equivalent optical path fluctuation $\delta x$ for
each passage of the light through the interferometer arm. For such
an optical path fluctuation, our single{-}arm interferometer would
suffer a phase shift
\begin{equation}\label{eqn6}
\delta \theta \left( t \right) \approx \frac{2\pi }{\lambda
}\delta x\left( t \right) \end{equation}
\noindent each way. As a
consequence, if the PSD of $\delta x$ is $S_{x}$, this noise
source would add an equivalent phase noise

\begin{equation}\label{eqn7}
S_{\delta \theta } \approx 2\frac{4\pi ^2}{\lambda ^2}S_x.
\end{equation}

In LISA the targeted sensitivity \cite{re01} requires that
$S_{\Delta F}^{1/2}(\omega)/m^2 \leq \sqrt{2} \cdot 3\times
10^{-15}\left( {\rm m}/{\rm s}^2 \right)/\sqrt{{\rm Hz}}$ for a
frequency $f$ above $f
> 0.1$ mHz. The corresponding requirement for the interferometer
is a path-length noise spectrum of $S^{1/2}_x \leq 20\,{\rm pm} /
\sqrt{{\rm Hz}}$. With these figure the noise in \eref{eqn5} and
that in \eref{eqn7} cross at $ \approx $ 3 mHz thus allowing to
relax the requirement for $\Delta F/m$ to:

\begin{eqnarray}\label{eqn8}
S_{\Delta F/m}^{1/2}&\leq& \sqrt{2}\times 3 \times 10^{-15}
\frac{{\rm m}}{{\rm s}^2 \sqrt{{\rm Hz}}}\sqrt{1+\left(
\frac{f}{3{\rm mHz}}\right)^4}
\nonumber \\
&\approx& 4.2\times 10^{ - 15}\frac{m}{s^2\sqrt {Hz} }\left[ {1 +
\left( {\frac{f}{3\,mHz}} \right)^2} \right]
\end{eqnarray}

The requirement in \eref{eqn8} needs to be qualified. Limiting the
noise in \eref{eqn4} by a requirement just for the differential
force noise becomes inaccurate at frequencies above some 3-4 mHz.
However, if the velocity fluctuations of the two proof{-}masses
are independent, this approach represents a \textit{worst case}
one. This is also the case for a partly correlated noise, provided
that correlation is assumed to work in the worst direction, i.e.
by mimicking differential motion.

The focus of the above discussion has been in term of coordinate
frames. One can however restate these performance requirements in
terms of coordinate independent, or gauge invariant quantities,
i.e. in term of the curvature tensor $R_{\mu \nu \sigma }^\lambda
$ only.

For a gravitational wave, the curvature tensor is equal, in the
Fourier domain, to

\begin{equation}\label{eqn9}
R_{\mu \nu \sigma }^\lambda = \frac{\omega ^2}{2c^2}\left(
{{\begin{array}{*{20}c}

\mathord{\buildrel{\lower3pt\hbox{$\scriptscriptstyle\leftrightarrow$}}\over
{0}} \hfill &
{\mathord{\buildrel{\lower3pt\hbox{$\scriptscriptstyle\leftrightarrow$}}\over
{H}} _ + } \hfill &
{\mathord{\buildrel{\lower3pt\hbox{$\scriptscriptstyle\leftrightarrow$}}\over
{H}} _\times } \hfill &
\mathord{\buildrel{\lower3pt\hbox{$\scriptscriptstyle\leftrightarrow$}}\over
{0}} \hfill \\

{\mathord{\buildrel{\lower3pt\hbox{$\scriptscriptstyle\leftrightarrow$}}\over
{H}} _ + } \hfill &
\mathord{\buildrel{\lower3pt\hbox{$\scriptscriptstyle\leftrightarrow$}}\over
{0}} \hfill &
\mathord{\buildrel{\lower3pt\hbox{$\scriptscriptstyle\leftrightarrow$}}\over
{0}} \hfill & { -
\mathord{\buildrel{\lower3pt\hbox{$\scriptscriptstyle\leftrightarrow$}}\over
{H}} _ + } \hfill \\

{\mathord{\buildrel{\lower3pt\hbox{$\scriptscriptstyle\leftrightarrow$}}\over
{H}} _\times } \hfill &
\mathord{\buildrel{\lower3pt\hbox{$\scriptscriptstyle\leftrightarrow$}}\over
{0}} \hfill &
\mathord{\buildrel{\lower3pt\hbox{$\scriptscriptstyle\leftrightarrow$}}\over
{0}} \hfill & { -
\mathord{\buildrel{\lower3pt\hbox{$\scriptscriptstyle\leftrightarrow$}}\over
{H}} _\times } \hfill \\

\mathord{\buildrel{\lower3pt\hbox{$\scriptscriptstyle\leftrightarrow$}}\over
{0}} \hfill &
{\mathord{\buildrel{\lower3pt\hbox{$\scriptscriptstyle\leftrightarrow$}}\over
{H}} _ + } \hfill &
{\mathord{\buildrel{\lower3pt\hbox{$\scriptscriptstyle\leftrightarrow$}}\over
{H}} _\times } \hfill &
\mathord{\buildrel{\lower3pt\hbox{$\scriptscriptstyle\leftrightarrow$}}\over
{0}} \hfill \\
\end{array} }} \right)
\end{equation}

\noindent with

$$\mathord{\buildrel{\lower3pt\hbox{$\scriptscriptstyle\leftrightarrow$}}\over
{H}} _ + = \left( {{\begin{array}{*{20}c}
 0 \hfill & { - h_ + \left( \omega \right)} \hfill & { - h_\times \left(
\omega \right)} \hfill & 0 \hfill \\
 {h_ + \left( \omega \right)} \hfill & 0 \hfill & 0 \hfill & { - h_ + \left(
\omega \right)} \hfill \\
 {h_\times \left( \omega \right)} \hfill & 0 \hfill & 0 \hfill & { -
h_\times \left( \omega \right)} \hfill \\
 0 \hfill & {h_ + \left( \omega \right)} \hfill & {h_\times \left( \omega
\right)} \hfill & 0 \hfill \\
\end{array} }} \right)$$
and
\begin{equation}\label{eqn10}
\mathord{\buildrel{\lower3pt\hbox{$\scriptscriptstyle\leftrightarrow$}}\over
{H}} _\times = \left( {{\begin{array}{*{20}c}
 0 \hfill & { - h_\times \left( \omega \right)} \hfill & {h_ + \left( \omega
\right)} \hfill & 0 \hfill \\
 {h_\times \left( \omega \right)} \hfill & 0 \hfill & 0 \hfill & { -
h_\times \left( \omega \right)} \hfill \\
 { - h_ + \left( \omega \right)} \hfill & 0 \hfill & 0 \hfill & {h_ + \left(
\omega \right)} \hfill \\
 0 \hfill & {h_\times \left( \omega \right)} \hfill & { - h_ + \left( \omega
\right)} \hfill & 0 \hfill \\
\end{array} }} \right) \end{equation}

For ${\left| \omega \right|L} \mathord{\left/ {\vphantom {{\left|
\omega \right|L} {2 \ll 1}}} \right. \kern-\nulldelimiterspace} {2
\ll 1}$ and for an optimally oriented ($\phi $=0) interferometer
arm the phase shift in \eref{eqn1} can then be written as:
\begin{equation}\label{eqn11}
\delta \theta \left( \omega \right) \approx \frac{2\pi }{\lambda
}Lh_ + \left( \omega \right) \approx \frac{2\pi }{\lambda
}2c^2L\frac{R\left( \omega \right)}{\omega ^2}
\end{equation}
\noindent with $R$ the generic component of the curvature tensor.

By comparing \eref{eqn11} with \eref{eqn8} $S_h^{1 \mathord{\left/
{\vphantom {1 2}} \right. \kern-\nulldelimiterspace} 2} \left(
\omega \right) \approx \frac{2}{\omega ^2L}S_{{\Delta F}
\mathord{\left/ {\vphantom {{\Delta F} m}} \right.
\kern-\nulldelimiterspace} m}^{1 \mathord{\left/ {\vphantom {1 2}}
\right. \kern-\nulldelimiterspace} 2} $ one can recast the
differential force noise as an effective curvature noise with
spectrum:

\begin{equation}\label{eqn12}
S_R^{1 \mathord{\left/ {\vphantom {1 2}} \right.
\kern-\nulldelimiterspace} 2} \left( \omega \right) \approx
\frac{1}{c^2L}S_{{\Delta F} \mathord{\left/ {\vphantom {{\Delta F}
m}} \right. \kern-\nulldelimiterspace} m}^{1 \mathord{\left/
{\vphantom {1 2}} \right. \kern-\nulldelimiterspace} 2}
\end{equation}

Thus to achieve its science goals, LISA must reach a curvature
resolution of order 10$^{-41}$ m$^{ - 2}$/$\surd $Hz or of
10$^{-43}$ m$^{-2 }$ for a signal at 0.1 mHz integrated over a
cycle. This figure may be compared with the scale of the curvature
tensor due to the gravitational field of the Sun at the LISA
location of $ \approx $ 10$^{-30}$ m$^{ - 2}$.

The aim of LISA Pathfinder mission of the European Space Agency
(ESA) is to demonstrate that indeed a TT frame may be constructed
by using particles nominally free orbiting within the solar
system, with accuracy relevant for LISA. Specifically within the
LISA Technology Package (LTP)\footnote{ The LTP is a collaboration
between ESA, and the space Agencies of Germany (DLR), Italy (ASI),
The Netherlands (SRON), Spain (MEC), Switzerland (SSO) and United
Kingdom (PPARC). In addition France (CNES/CNRS) is in the process
of joining the team.}, two LISA-like proof{-}masses located inside
a single spacecraft are tracked by a laser interferometer. This
minimal instrument is deemed to contain the essence of the
construction procedure needed for LISA and thus to demonstrate its
feasibility. This demonstration requires two steps:
\begin{itemize}
\item Firstly, based on a noise model \cite{re04,re05}, the
mission is designed so that any differential parasitic
acceleration noise of the proof{-}masses is kept below the
requirements. For the LTP these requirements are relaxed to
$3\times 10^{ - 14}{\rm m}{\rm s}^{ - 2} / \sqrt{ {\rm Hz}}$, a
factor $ \approx $ 7 larger than what is required in LISA. In
addition this performance is only required for frequencies larger
than 1 mHz. This relaxation of performance is accepted in view of
cost and time saving.

As both for LISA and for the LTP this level of performance cannot
be verified on ground due to the presence of the large Earth
gravity, the verification is mostly relying on the measurements of
key parameters of the noise model of the instrument
\cite{re06,re07,re08,re09,re10}. In addition an upper limit to all
parasitic forces that act at the proof-mass surface
(electrostatics and electromagnetics, thermal and pressure effects
etc.) has been established and keeps being updated by means of a
torsion pendulum test{-}bench \cite{re07,re11,re12}. In this
instrument a hollow version of the proof{-}mass hangs from the
torsion fiber of the pendulum so that it can freely move in a
horizontal plane within a housing which is representative of
flight conditions. An equivalent differential acceleration noise
of $\approx 3\times 10^{ - 13}{\rm m}{\rm s}^{ - 2} / \sqrt{ {\rm
Hz}}$ has been measured \cite{re07}. \item Secondly, once in orbit
the residual differential acceleration noise of the proof{-}masses
is measured. The noise model \cite{re13,re14} predicts that the
contributions to the total PSD fall into three broad categories:
\begin{itemize}
\item Noise sources whose effect can be identified and suppressed
by a proper adjustment of selected instrument parameters. An
example of this is the force due to residual coupling of
proof{-}masses to the spacecraft. By regulating and eventually
matching, throughout the application of electric field, the
stiffness of this coupling for both proof{-}masses, this source of
noise can be first highlighted, then measured, and eventually
suppressed \cite{re14}.

\item Noise sources connected to measurable fluctuations of some
physical parameter. Forces due to magnetic fields or to thermal
gradients are typical examples. The transfer function between
these fluctuations and the corresponding differential proof{-}mass
acceleration fluctuations will be measured by purposely enhancing
the variation of the physical parameter under investigation
\cite{re14} and by measuring the corresponding acceleration
response. For instance the LTP carries magnetic coils to apply
comparatively large magnetic field signals and heaters to induce
time{-}varying thermal gradients. In addition the LTP also carries
sensors to measure the fluctuation of the above physical
disturbances while measuring the residual differential
acceleration noise in the absence of any applied perturbation.
Examples of these sensors are magnetometers and thermometers to
continue with the examples above. By multiplying the measured
transfer function by the measured disturbance fluctuations, one
can generate an expected acceleration noise data stream to be
subtracted from the main differential acceleration data stream.
This way the contributions of these noise sources are suppressed
and the residual acceleration PSD decreased. This reduction is
obtained without requiring expensive magnetic ``cleanliness'', or
thermal stabilization programs.

\item Noise sources that cannot be removed by any of the above
methods. The residual differential acceleration noise due to these
sources must be accounted for \cite{re14}. To be able to do the
required comparison, some of the noise model parameters must and
will be measured in{-}flight. One example for all, the charged
particle flux due to cosmic rays will be continuously monitored by
a particle detector.
\end{itemize}
\end{itemize}
The result of the above procedure is the validation of the noise
model for LISA and the demonstration that no unforeseen source of
disturbance is present that exceeds the residual uncertainty on
the measured PSD. The following sections, after describing some
details of the experiment, will discuss the expected amount of
this residual uncertainty.

\section{The LISA Technology Package experiment}
\label{sec:mylabel1}

The basic scheme of the LTP has been described in \cite{re02} and
is shown in \fref{fig1}:
\begin{figure}[t]
\centerline{\includegraphics[height=8.00cm]{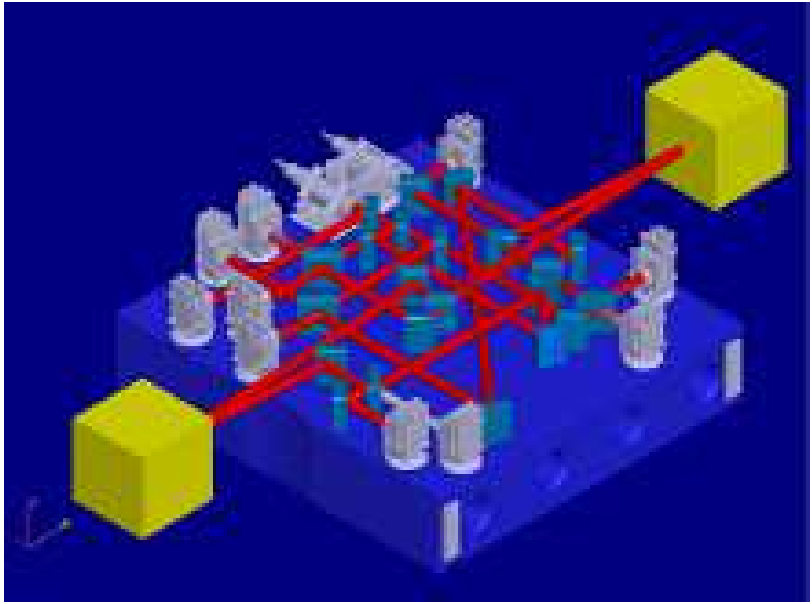}}
\caption{The concept of the LTP: The distance between 2 cubic,
free floating proof{-}masses is measured by a heterodyne laser
interferometer. The proof{-}masses centers nominal separation is
L=376 mm, the proof{-}mass size is 46 mm and the mass value is
1.96 kg.} \label{fig1}
\end{figure}
\noindent two free{-}floating proof{-}masses are hosted within a
single spacecraft and the relative motion along a common sensitive
axis, the {\bf x}-axis, is measured by means of a laser
interferometer. The proof{-}masses are made of a Gold-Platinum,
low magnetic susceptibility alloy, have a mass of $m= 1.96$ kg and
are separated by a nominal distance of 376 mm. A picture of the
instrument in its current stage of design and a few pictures of a
prototype under development are reported from \fref{fig2} to
\fref{fig4}. In the following section we summarize the basic
elements.

\subsection{The LTP instrument}

In the LTP, as in LISA, each proof{-}mass is surrounded by a set
of electrodes that are used to readout the mass position and
orientation relative to the spacecraft \cite{re11,re12,re15,re16}
(\fref{fig3}). This measurement is obtained as the motion of the
proof{-}mass varies the capacitances between the electrodes and
the proof{-}mass itself. The same set of electrodes is also used
to apply electrostatic forces to the proof{-}masses.

\begin{figure}[t]
\centerline{
\includegraphics[height=9cm]{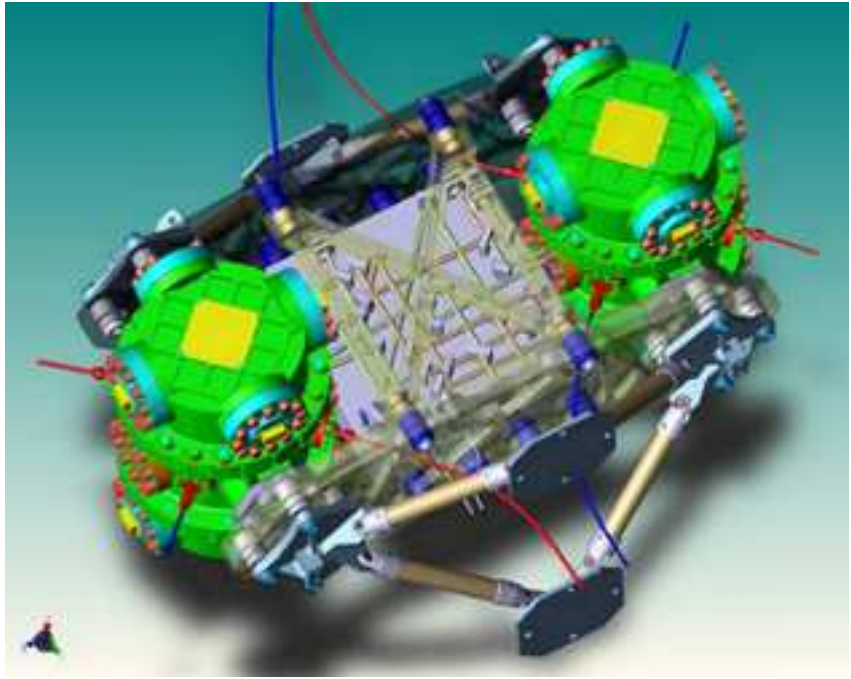}} \caption{The LISA Technology Package. The green chambers are the
vacuum enclosures that provide autonomous ultra{-}high vacuum
around the proof-masses (non visible). An optical bench in between
the proof{-}masses (in grey) supports the interferometry that
reads out the distance between the masses. The interferometer
laser beam hits each proof{-}mass by crossing the vacuum
enclosures through an optical window. The entire supporting
structure is made out of glass-ceramics for high
thermo{-}mechanical stability. The red and blue lines are the
optical fibers that carry the UV light used for contact{-}less
discharging of proof{-}masses. Also visible are the side struts
that connect the LTP to the spacecraft} \label{fig2}
\end{figure}

Differential capacitance variations are parametrically read out by
a front{-}end electronics composed of high accuracy differential
inductive bridges excited at about 100 kHz, and synchronously
detected via a phase sensitive detector \cite{re15,re17}.
Sensitivities depend on the degree of freedom. For the {\bf
x}-axis it is better than 1.8 nm/$\sqrt{{\rm Hz}}$ at 1 mHz.
Angular sensitivities are better than 200 nrad/$\sqrt{{\rm Hz}}$.

Forces and torques on the proof{-}masses required during science
operation are applied through the same front{-}end electronics by
modulating the amplitude of an ac carrier applied to the
electrodes. The frequency of the carrier is high enough to prevent
the application of unwanted forces by mixing with low frequency
fluctuating random voltages. The front end electronics is also
used to apply all voltages required by specific experiments
\cite{re17}.

\begin{figure}[t]
\centerline{\includegraphics[width=9.56cm,height=7.50cm]{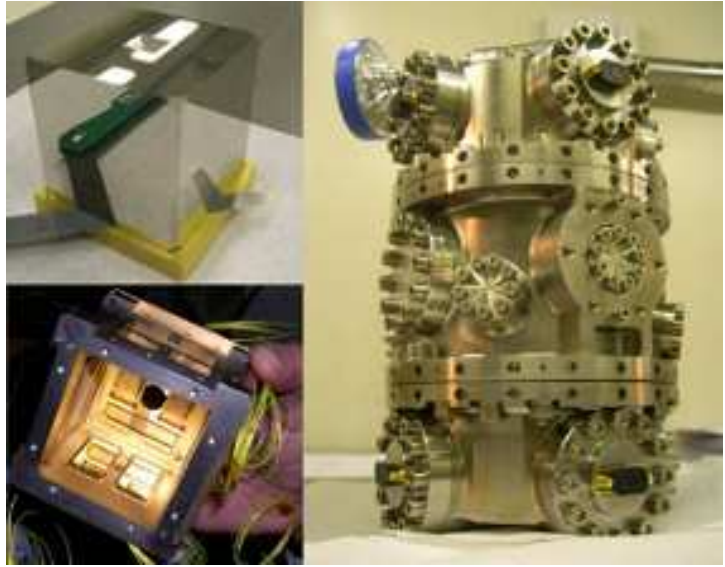}}
\caption{A prototype of the gravity reference sensor under
development. Top left: the gold{-}platinum proof{-}mass. Bottom
left: the electrode housing carrying the Gold{-}coated ceramics
electrodes. Right: the vacuum enclosure. The optical windows for
the laser light are substituted by plane flanges.} \label{fig3}
\end{figure}

\begin{center}
\begin{figure}[t]
\centerline
{\includegraphics[width=5.6cm,height=5cm]{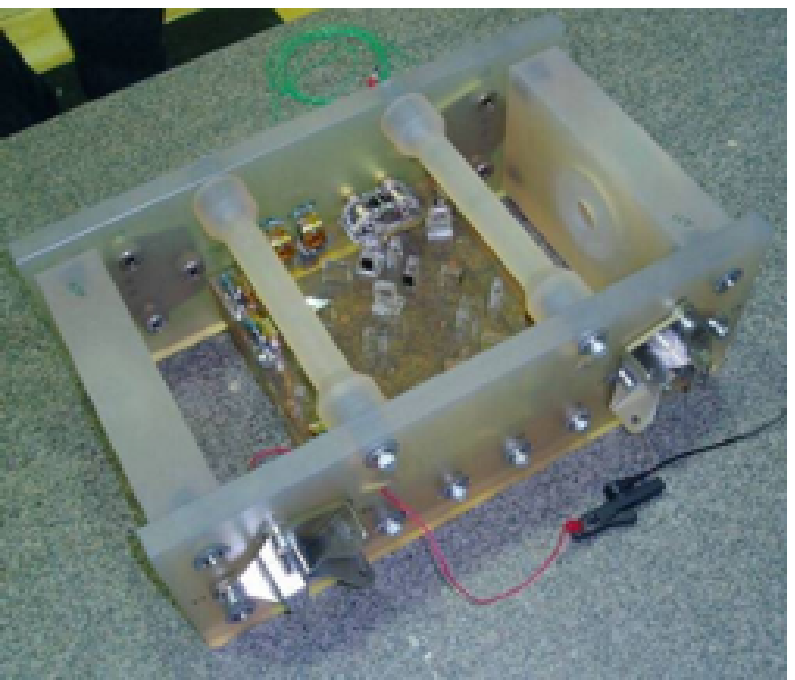}
\includegraphics[width=6.66cm,height=5cm]{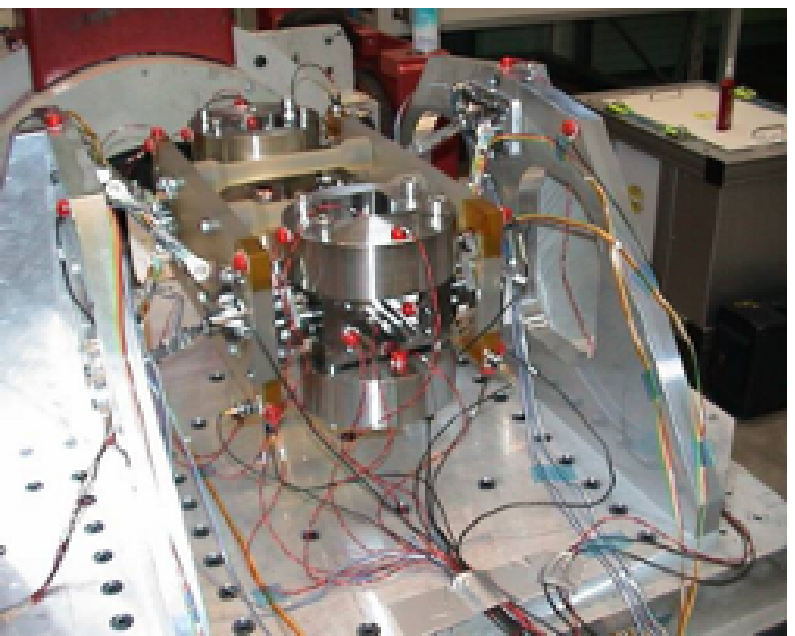}}
\caption{Left: a prototype of the optical bench under development.
Right: vibration test of the optical bench with two dummies of the
gravity reference sensors to simulate the launch loads.}
\label{fig4}
\end{figure}
\end{center}

\begin{figure}[htbp]
\centerline{\centerline
{\includegraphics[width=13cm]{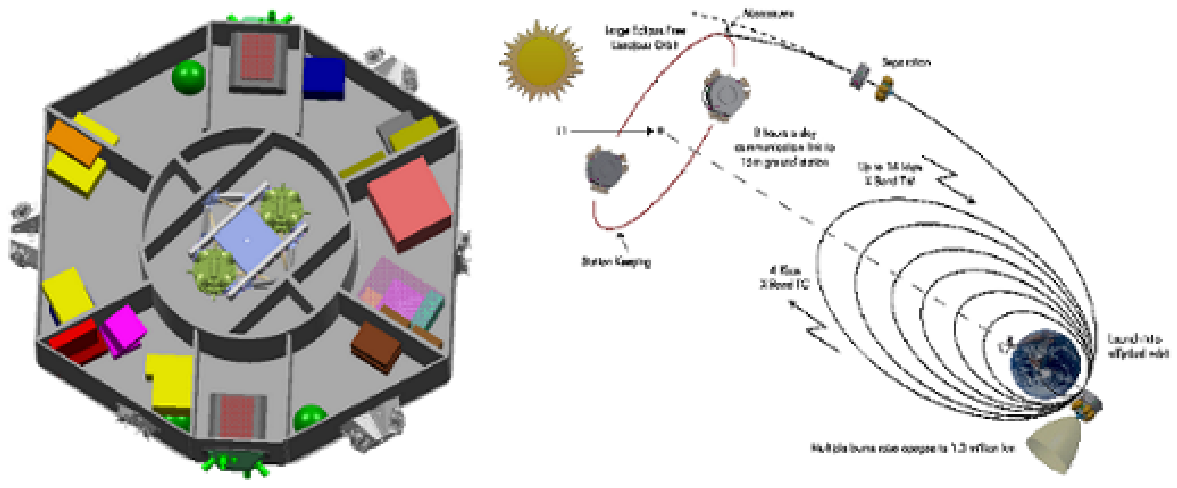}}} \caption{Left:
the LTP accommodated within the central section of the LISA
Pathfinder spacecraft. Right: the injection of LISA Pathfinder in
the final orbit around L1.} \label{fig5}
\end{figure}

Each proof{-}mass, with its own electrode housing, is enclosed in
a high vacuum chamber which is pumped down to 10$^{ - 5}$ Pa by a
set of getter pumps. The laser interferometer light passes through
vacuum chamber wall through an optical window.

As the proof{-}mass has no mechanical contact to its surrounding,
its electrical charge continues to build up due to cosmic rays. To
discharge the proof{-}mass, an ultra{-}violet light is shone on it
and/or on the surrounding electrodes \cite{re18}. Depending on the
illumination scheme, the generated photoelectrons can be deposited
on or taken out of the proof{-}mass to achieve electrical
neutrality.

The absence of a mechanical contact also requires that a blocking
mechanism keeps the mass fixed during launch and is able to
release it overcoming the residual adhesion. This release must
leave the proof{-}mass with small enough linear momentum to allow
the control system described in the following to bring it at rest
in the nominal operational working point.

The system formed by one proof{-}mass, its electrode housing, the
vacuum enclosure and the other subsystems is called in the
following the gravity reference sensor (GRS).

The interferometer system provides the following measurements: 1)
heterodyne measurement of the relative position of the
proof{-}masses along the sensitive axis. 2) Heterodyne measurement
of the position of one of the proof-masses (proof{-}mass 1)
relative to the optical bench. 3) Differential wave{-}front
sensing of the relative orientations of the proof-masses around
the {\bf y}-axis and the {\bf z}-axis. 4) Differential
wave{-}front sensing of the orientation of proof-mass 1 around the
{\bf y}-axis and {\bf z}{-}axis. Sensitivities at mHz frequency
are in the range of 10 pm/$\sqrt{{\rm Hz}}$ for displacement and
of 10 nrad/$\sqrt{{\rm Hz}}$ in rotation.

As a light source for the heterodyne interferometry, a
diode-pumped, monolithic Nd:YAG non-planar ring oscillator is
used. To obtain the necessary frequency shift, the beam coming
from the laser is split and each partial beam is sent through an
acousto-optical modulator (AOM). The light is then delivered to
the optical bench by a pair of optical fibres and fibre injectors.
Quadrant photo-diodes are used for the detection of the
interferometric signal, allowing to measure yaw and pitch of the
proof-masses with respect to the sensitive axis. The optical
components needed for the interferometer (i.e. mirrors and
beam{-}splitters) are attached to the optical bench through
hydroxyl-silicate bonding to ensure mechanical stability.

Data acquisition, conditioning and phase-measurement is performed
by the interferometer front-end electronics, based largely on
Field Programmable Gate Arrays (FPGA). The final processing and
retrieval of the position signals from the phase measurements is
performed by the LTP payload computer. A full description of the
interferometer for can be found in \cite{re19}.

The LTP computer also drives and reads-out the set of subsidiary
sensor and actuators \cite{re20} needed to apply the already
mentioned selected perturbations to the proof{-}masses and to
measure the fluctuations of the disturbing fields. Actuators
include coils used to generate magnetic field and magnetic field
gradients and heaters to vary temperature and temperature
differences at selected points of the Gravity Reference Sensor and
of the optical bench. Sensors include magnetometers, thermometers,
particle detectors and monitors for the voltage stability of the
electrical supplies.

The LTP will fly on LISA Pathfinder \cite{re21} in 2008. It will
be hosted in the central section of the spacecraft (\fref{fig5},
left), where gravitational disturbances are minimized, and will
operate in a Lissajous orbit \cite{re22} around the Lagrange point
1 of the Sun-Earth system, an environment very similar to that
where the LISA spacecrafts will operate (\fref{fig5}, right).
Further details can be found in \cite{re21}.

\subsection{Experiment performance: similarities and differences with LISA.}

Within the LTP, as in LISA, a key element for suppressing the
force disturbance is that the proof{-}masses have no mechanical
contact to the spacecraft. In addition, as forces may depend on
the position of the proof{-}masses within the spacecraft, this is
kept as fixed as possible.

To fulfill both of these apparently conflicting requirements,
\cite{re01} the spacecraft actively follows the proof{-}mass
located within it, in a closed loop control scheme usually called
the drag{-}free control. The position of the proof{-}mass relative
to some nominal origin is measured by means of the gravitational
reference sensor. A high gain control loop tries to null this
error signal by forcing the spacecraft to follow the proof{-}mass.
In order to produce the necessary force on the spacecraft, the
control loop drives a set of micro{-}thrusters.

With this control scheme, if the loop gain is high enough, the
difference of force between two masses sitting in two different
spacecraft can be calculated \cite{re02} to be:
\begin{equation}\label{eqn13}
\frac{\Delta F}{m} = \frac{\Delta f}{m} - \omega _{p2}^2 \delta
x_2 + \omega _{p1}^2 \delta x_1 \end{equation} Here $\Delta $f is
the difference of position{-}independent, fluctuating forces
directly acting on proof{-}masses, $\omega _{p1(2)}^2 $ is the
stiffness per unit mass of parasitic spring coupling of
proof{-}mass 1(2) to the spacecraft and $\delta x_{1(2)} $is the
residual jitter of the same proof-mass relative to the spacecraft.

The LTP experiment uses a similar drag{-}free control scheme.
However here both masses sit in one spacecraft and cannot be
simultaneously followed by it. One of them must be forced to
follow either the other one or the spacecraft, possibly within a
closed loop control scheme. The small required force is applied
via the capacitive actuation described above and this control loop
is usually called the electrostatic suspension.

Various modes of control are envisaged \cite{re23,re24}. In the
basic one the spacecraft follows one of the proof{-}masses, say
n.1, along the direction of the interferometer arm. In addition,
the main interferometer readout, i.e. the distance between the
proof{-}masses, is used as the error signal to actuate the second
proof{-}mass along the same axis. This way the distance between
the proof{-}masses is kept constant.

The laser interferometer output, at angular frequency $\omega $,
is given, in the Fourier space, by:
\begin{equation}\label{eqn14}
\delta x_{{\rm ifo}} = x_{n,{\rm ifo}} + \frac{\Delta F}{\omega
^2m}
\end{equation}
\noindent where $x_{n,{\rm ifo}}$ is the interferometer
displacement noise and the difference of force on the right hand
side includes all possible contributions.

$\Delta F/m$ is affected by the control loops. In the
approximation of high drag{-}free gain and small parasitic
stiffness, $\Delta F/m$ is given by \cite{re14}:
\begin{equation}\label{eqn15}
\frac{\Delta F}{m} \approx \frac{\omega ^2}{\omega ^2 - \omega
_{{\rm es}}^2 \left( \omega \right)}\left\{ {\frac{\Delta f}{m} +
\left( {\omega _{p1}^2 - \omega _{p2}^2 } \right)\delta x_1 +
\omega _{{\rm es}}^2 \left( \omega \right)x_{n,{\rm ifo}} }
\right\}
\end{equation} \noindent where $\omega _{{\rm es}}^2 \left( \omega
\right)$ is the gain (per unit mass) of the electrostatic
suspension loop. $\Delta f$ in the parentheses is again the
position{-}independent difference of forces that would act on the
proof-masses \textit{in the absence of any control loop}.

The effective measurement of $\Delta F/m$ is performed by
measuring the signal $\delta x_{ifo} $, or even better its second
time derivative. By using \eref{eqn15} one gets:
\begin{equation}\label{eqn16}
\omega ^2\delta x_{{\rm ifo}} \approx \frac{\omega ^2}{\omega ^2 -
\omega _{{\rm es}}^2 \left( \omega \right)}\left\{ {\frac{\Delta
f}{m} + \left( {\omega _{p1}^2 - \omega _{p2}^2 } \right)\delta
x_1 + \omega ^2x_{n,{\rm ifo}} } \right\} \end{equation}

\Eref{eqn16} brings a remarkable similarity to \eref{eqn13}, thus
suggesting that indeed the basic concept of the experiment is
sound. However there are a few remarks that need to be done.

In the LTP both proof{-}masses are spring{-}coupled to the same
spacecraft and both feel then the relative jitter between this one
and the drag{-}free reference proof{-}mass 1. Some care must then
be used to avoid that, by getting $\omega _{p1}^2 \simeq \omega
_{p2}^2 $ a substantial residual jitter $\delta x_{1}$ becomes
unobservable, thus yielding a too small estimate of the noise in
\eref{eqn13}. This is easily avoided by a detailed sequence of
measurements for both $\omega _{p1}^2 $ and $\omega _{p2}^2$ that
has been described in \cite{re14}.

Correlation of disturbances on different proof-masses may play a
different role in LISA and in the LTP. In LISA proof-masses within
the same interferometer arm belong to different spacecraft and are
located 5$\times $10$^{9}$~m apart. The only correlated
disturbances one can think of are connected to the coupling to the
Sun: magnetic field fluctuations, fluctuation of the flux of
charged particles in solar flares and the fluctuation of solar
radiation intensity that may induce correlated thermal
fluctuations in distant spacecraft. These correlations will only
slightly affect the error budget and will have no profound
consequences on the experiment itself.

In the LTP all sources of noise that share the same source for
both proof{-}masses are correlated. Magnetic noise generated on
board the spacecraft, thermal fluctuations and gravity noise due
to thermo{-}elastic distortion of spacecraft constitute a few
examples. The major concern with correlated noise is that, by
affecting the proof-masses the same way, it might subtract from
the differential measurement. This subtraction would not happen in
LISA and thus would cause an optimistic underestimate of the total
noise. This possibility can reasonably be avoided for almost all
candidate effects: magnetic disturbances are quadratic function of
the field and can then be modulated by purposely applying
asymmetric fields and gradient to different proof{-}masses (that
will anyhow have susceptibility and remnant moments only matched
not better than 50{\%}). Sensitivity to thermal gradient can be
modulated by independently adjusting the static temperature of
each proof{-}mass. Similar strategies of intentional mismatch of
the response function of the two proof{-}masses can be devised for
almost all disturbances.

An exception is constituted by the gravitational noise for which
the response, due to the equivalence principle, cannot be changed.
However realistic assumptions about thermal distortion make the
event of a gravity fluctuation affecting both proof-masses with
the same force along the {\bf x}-axis very unlikely.

In addition LISA Pathfinder also carries the independent NASA
provided ST-7 technology package \cite{re25}, based on the same
concept as the LTP that will operate for part of the mission
jointly with the LTP. The possibility that these gravitational
disturbances cancel exactly and simultaneously on both instruments
is even more unlikely if not impossible. Notice that the joint
operation between LTP and ST-7 will also allow to detect some of
the correlated disturbances discussed above.

The presence of the electrostatic suspension multiplies the forces
in \eref{eqn16} by the transfer function $\omega ^2\left[ {\omega
^2 - \omega _{{\rm es}}^2 \left( \omega \right)} \right]^{ - 1}$.
Though the numerical algorithm behind $\omega _{{\rm es}}^2 \left(
\omega \right)$ is entirely known, this gain also includes the
conversion from a force command issued by a computer to actual
voltages and forces on the proof{-}masses, thus posing a
calibration issue if $\big| \omega_{{\rm es}}^2(\omega) \big|
\succsim \omega^2$. If instead $\left| {\omega _{{\rm es}}^2
\left( \omega \right)} \right| \ll \omega ^2$, then the $\omega
^{2}\delta x_{{\rm ifo}}$ is a faithful representation of the
force signal. This last limit is then accurate and can be used for
calibration, but it brings about the disadvantage that $\omega
_{{\rm es}}^2 \left( \omega \right)$ must be kept small at all
relevant frequencies, thus producing long and slightly unpractical
relaxation time constants in the electrostatic suspension loop. As
an alternative, within the joint operation with ST-7, a
calibration procedure is envisaged where a comparatively large
motion of one of the proof-masses of the ST-7 package is used to
generate an oscillating gravitational field on the LTP
proof-masses. This will allow an independent absolute calibration
of forces on board at better than 1{\%}.

Finally \eref{eqn16} shows that the readout noise contributes the
feedback force $\omega ^2x_{n,{\rm ifo}} $ that is not present in
the case of LISA. This is discussed in the next section.

\subsection{Experiment performance: sensitivity}
\label{subsec:experiment}

As an instrument to measure $\Delta F/m$, the LTP is then limited
by the laser interferometer noise. When $\left| {\omega _{{\rm
es}}^2 \left( \omega \right)} \right| \ll \omega ^2$ the signal
is:
\begin{equation}\label{eqn17}
\omega ^2\delta x_{ifo} \approx \frac{\Delta f}{m} + \left(
{\omega _{p1}^2 - \omega _{p2}^2 } \right)\delta x_1 + \omega
^2x_{n,{\rm ifo}} \end{equation}

\Eref{eqn18} shows that, as in LISA, the laser interferometer
noise converts into an effective force noise according to

\begin{equation}\label{eqn18}
S_{{\Delta F} \mathord{\left/ {\vphantom {{\Delta F} m}} \right.
\kern-\nulldelimiterspace} m}^{1 \mathord{\left/ {\vphantom {1 2}}
\right. \kern-\nulldelimiterspace} 2} \left( \omega \right)
\approx \omega ^2S_{n,{\rm ifo}}^{1 \mathord{\left/ {\vphantom {1
2}} \right. \kern-\nulldelimiterspace} 2} \left( \omega \right)
\end{equation}

For the LTP the laser interferometer is requested to achieve a
sensitivity of:
\begin{equation}\label{eqn19}
S_{n,{\rm laser}} =\left( 9 {\rm pm} / \sqrt{{\rm
Hz}}\right)\sqrt{1+ \left(\frac{\omega}{2\pi\times 3 {\rm mHz}}
\right)^{-4}}
\end{equation}

This noise corresponds to an equivalent force noise of:
\begin{equation}\label{eqn20}
S_{{\Delta F} \mathord{\left/ {\vphantom {{\Delta F} m}} \right.
\kern-\nulldelimiterspace} m}^{1 \mathord{\left/ {\vphantom {1 2}}
\right. \kern-\nulldelimiterspace} 2} \left( f \right) \approx
\left( {{3.2\times 10^{ - 15}{\rm m}{\rm s}^{ - 2}}
\mathord{\left/ {\vphantom {{3.2\times 10^{ - 15}ms^{ - 2}} {\sqrt
{{\rm Hz}} }}} \right. \kern-\nulldelimiterspace} {\sqrt {{\rm
Hz}} }} \right)\left[ {1 + \left( {\frac{f}{3{\rm mHz}}}
\right)^2} \right] \end{equation}

At lower frequencies an additional force noise adds up to mask the
parasitic forces due to other sources. This force noise is due to
the fluctuations of the gain of the electrostatic suspension loop.
Indeed the electrostatic suspension must also cope with any static
force acting on the proof{-}masses. If the force stays constant
but the gain fluctuates, the feedback force fluctuates
consequently, adding a noise source that is expected to limit the
sensitivity at the lowest frequencies. This effect only appears in
the LTP as in LISA static forces are compensated just by the
drag-free loop, and no electrostatic suspension is envisaged. The
largest expected source of gain fluctuations is the fluctuation of
the dc voltage which is used to stabilize the actuation
electronics.

A relative fluctuation of the voltage $\delta $V/V would then
produce a fluctuation of force:
\begin{equation}\label{eqn21}
\frac{\delta f}{m} \approx \left( {\frac{\Delta f}{m}}
\right)_{{\rm static}} \times 2\frac{\delta V}{V} \end{equation}
\noindent where the factor 2 comes from the quadratic conversion
from voltage to force.

\begin{figure}[htbp]
\centerline{\includegraphics[width=13cm,height=7.80cm]{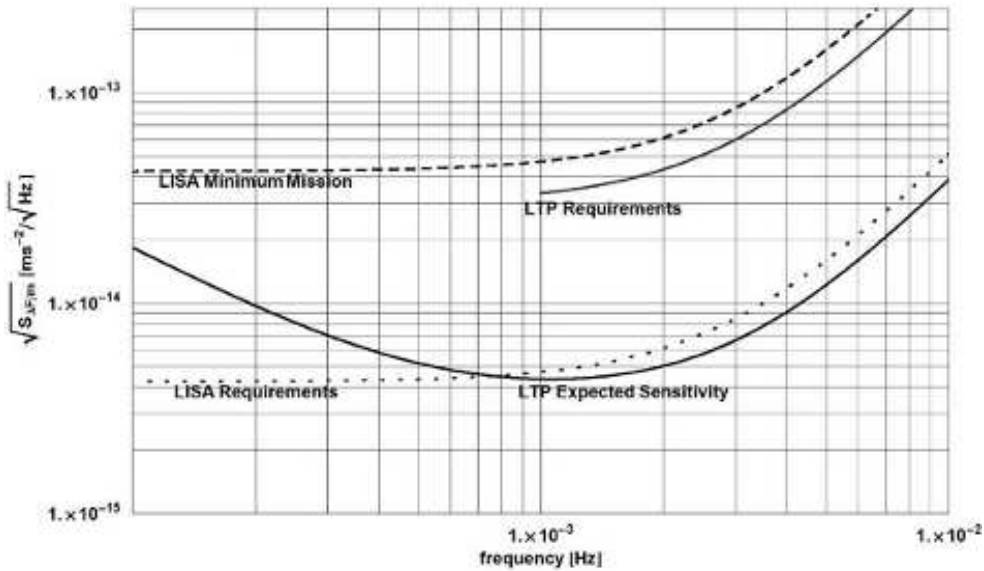}}
\caption{Lower solid curve: projected sensitivity for differential
force measurement of the LTP experiment. Upper solid curve:
required maximum differential acceleration noise for the LTP.
Lower dotted curve:, LISA requirements from \eref{eqn8}. Upper,
dashed curve: LISA Minimum mission requirements.} \label{fig6}
\end{figure}

For the LTP it is required that $\left| {{\Delta f}
\mathord{\left/ {\vphantom {{\Delta f} m}} \right.
\kern-\nulldelimiterspace} m} \right|_{{\rm static}} \le 1.3\times
10^{ - 9}{\rm m}{\rm s}^{ - 2}$. However to keep some margin,
balancing of the gravitational field, the largest source of dc
forces, is requested at one half of the above figure and current
models predict that the remaining sources of dc forces are
negligible. In addition the dc-voltage reference is required to be
stable to within $S_{{\delta V} \mathord{\left/ {\vphantom
{{\delta V} V}} \right. \kern-\nulldelimiterspace} V}^{1
\mathord{\left/ {\vphantom {1 2}} \right.
\kern-\nulldelimiterspace} 2} \le 2\times 10^{ - 6}1
\mathord{\left/ {\vphantom {1 {\sqrt {{\rm Hz}} }}} \right.
\kern-\nulldelimiterspace} {\sqrt {Hz} }$ at 1 mHz. As a goal this
stability should degrade not worse than 1/f at lower frequencies
down to 0.1 mHz. If one can achieve these goals, gain fluctuations
may convert to an equivalent force PSD of
\begin{equation}\label{eqn22} S_{{f_{fb} } \mathord{\left/
{\vphantom {{f_{{\rm fb}} } m}} \right. \kern-\nulldelimiterspace}
m}^{1 \mathord{\left/ {\vphantom {1 2}} \right.
\kern-\nulldelimiterspace} 2} \left( f \right) \approx 1.8\times
10^{ - 15}\frac{ms^{ - 2}}{\sqrt {Hz} }\sqrt {1 + \left(
{\frac{1mHz}{f}} \right)^2}. \end{equation} Adding up the results
in \eref{eqn22} and \eref{eqn20} one gets the sensitivity
prediction in \fref{fig6}.

In \fref{fig6} we also report for comparison, besides the LISA
requirements from \eref{eqn8} and the LTP maximum noise
requirements, the required sensitivity for mission success for
LISA, the so called ``minimum mission'' requirements. This mission
would still detect gravitational waves from merger of black holes
with 3 x 10$^{5}$ M$_{\varepsilon }$ at $z=1$ with high S/N ratio.
Furthermore it would be able to detect and study the waves from
thousands of Galactic compact binaries. Finally it would detect at
SNR $>$ 10 one or more of the well identified galactic binaries
that are usually called ``verification binaries'' as everything is
known about the expected gravitational wave signal.

\section{Concluding remarks}

\Fref{fig6} shows that the ultimate uncertainty on the
differential acceleration PSD can be potentially constrained by
the LISA Pathfinder mission below a factor 5 above LISA
requirements at 0.1 mHz, and near LISA requirements at 1 mHz or
above.

In addition, within the entire frequency range, the LISA
Pathfinder mission will constrain the acceleration noise somewhere
in the range between 1 and 10 fm~s$^{ - 2} /\sqrt{{\rm Hz}}$ well
below the requirements of LISA minimum mission, thus strongly
reducing the risk of a LISA failure.

Notice that the resulting TT frame, a frame where free particles
at rest remain at rest, is a very close approximation to the
classical concept of inertial frame, and would indeed be inertial,
within the measurement bandwidth, wouldn't it be for the presence
of the gravitational wave. Thus LISA Pathfinder will demonstrate
the possibility of building an inertial frame in a standard
spacecraft orbiting the Sun on a scale of a meter in space and of
a few hours in time at the above mentioned level of absence of
spurious accelerations. To our knowledge this will be at least 2
orders of magnitude better than what will be achieved by the GOCE
mission \cite{re26} which will in turn be much better than any
existing current limit.


\section*{References}
\begin{thebibliography}{99}

\bibitem{re01} Bender P. et al., ``LISA Pre-Phase-A report'' MPQ 208 (Munich
1996, unpublished).

\bibitem{re02} Vitale S., et al Nuclear Physics B (Proc. Suppl.)\textbf{ 110},
209 (2002)

\bibitem{re03} Misner C. W. , Thorne K. S., and Wheeler J. A. ``Gravitation''
(Freeman {\&} Co., San Francisco, 1973)

\bibitem{re04} Stebbins R. T., et al, Class. Quantum Grav. \textbf{21} (2004)
S653--S660

\bibitem{re05} Vitale S 2002 LISA Technology Package Architect Final Report ESTEC
contract no 15580/01/NL/HB

\bibitem{re06} Hueller M., et al., Class. Quantum Grav. \textbf{19} (2002)
1757--1765

\bibitem{re07} Carbone L., et al 2003 Phys. Rev. Lett. \textbf{91} 151101

\bibitem{re08} Carbone L. et al., Class. Quantum Grav. \textbf{21} (2004)
S611--S620

\bibitem{re09} Carbone L. et al., Class. Quantum Grav. (this issue) to appear
(2005); also gr-qc/0412103

\bibitem{re10} Hueller M. et al., Class. Quantum Grav. (this issue) to appear
(2005); also gr-qc/0412093

\bibitem{re11} Vitale S., and Dolesi R., AIP Conference Proceedings \textbf{523},
231 (2000)

\bibitem{re12} Cavalleri A., et al., Class. Quantum Grav. 18 (2001) 4133--4144

\bibitem{re13} Bortoluzzi D., et al., Class. Quantum Grav. \textbf{20} (2003)
S89--S97

\bibitem{re14} Bortoluzzi D., et al, Class. Quantum Grav. \textbf{21} (2004) S573

\bibitem{re15} Weber W. J., et al., SPIE Proc. \textbf{4856}, 31 (2002)

\bibitem{re16} Dolesi R., et al., Class. Quantum Grav. \textbf{20} (2003)
S99--S108

\bibitem{re17} Mance D., Zweifel, P., ``LISA Pathfinder Mission LTP - Inertial
Sensor Front-End Electronics'' S2-ETH-RS-3001, (ETH- Zurich 2004)

\bibitem{re18} Sumner T., et al., Class. Quantum Grav. \textbf{21} (2004)
S597--S602

\bibitem{re19} Heinzel G., et al., Class. Quantum Grav. 21 (2004) S581--S587

\bibitem{re20} Lobo A. et al., ``Data Diagnostics System science requirements
document'', S2-IEEC-RS-3002, (IEEC, Barcelona 2004)

\bibitem{re21} Dunbar N., Tomkins K., Gould K., Lecuyot A., Holt T., Fichter W.,
``LISA Pathfinder system design synthesis report'', S2.ASU.RP.2003
(EADS-ASTRIUM 2004)

\bibitem{re22} Landgraf M., Heckler M., and Kemble S., Class. Quantum Grav. (this
issue) to appear (2005)

\bibitem{re23} Bortoluzzi D., et al, Class. Quantum Grav. \textbf{20} (2003)
S227--S238

\bibitem{re24} Fichter W., et al., Quantum Grav. (this issue) to appear (2005)

\bibitem{re25} http://nmp.jpl.nasa.gov/st7/

\bibitem{re26} http://www.esa.int/export/esaLP/goce.html

\end{thebibliography}
\end{document}